\documentclass[twocolumn,longbib]{aastex7}

\usepackage{amsmath}
\usepackage[cm]{sfmath}

\newcommand{\mearth}{$\mathrm{M}_\oplus$}
\newcommand{\degrees}{$^{\circ}$}

\received{21 May 2026}
\revised{6 July 2026}
\accepted{7 July 2026}

\submitjournal{ApJL}

\usepackage[legacy]{csvsimple}
\usepackage{longtable}
\usepackage{postnotes}
\postnotesetup{format={\small},listenv=none,heading={{\bf\normalsize References. }},pretextmark={(},posttextmark={) } ,postprintnote={;  },makemark = {#2\hbox{{\small\normalfont#1}}#3} ,maketextmark = {#2#1#3}}

\usepackage{aas_macros}

\shorttitle{Growing Planets Produce Extreme Dust Signatures}
\shortauthors{Carter \& Leinhardt}

\begin{document}

\title{Growing Planets Produce Extreme Dust Signatures}

\author[orcid=0000-0001-5065-4625,gname=Philip,sname=Carter]{Philip J. Carter}
\affiliation{School of Physics, H.H. Wills Physics Laboratory, University of Bristol, Bristol BS8 1TL, UK}
\email[show]{p.carter@bristol.ac.uk}

\author[orcid=0000-0003-4813-7922]{Zo\"{e} M. Leinhardt}
\affiliation{School of Physics, H.H. Wills Physics Laboratory, University of Bristol, Bristol BS8 1TL, UK}
\email[show]{zoe.leinhardt@bristol.ac.uk}

\correspondingauthor{Philip J. Carter}

%%%%%

\begin{abstract}

\noindent Collisions between planetary bodies are essential to the assembly of rocky planets like the Earth, but they are extremely difficult to observe. 
We therefore rely on indirect signatures of planetary impacts such as extreme debris disks -- bright infrared excesses around other stars. Extreme debris disks are thought to be the result of energetic collisions that ejected large amounts of vaporized rock which rapidly condensed into mm- or cm-sized dust. 
However, previously there has been no clear way to relate the observed mass of dust to the collision that produced it. 
Here, we show that the colliding bodies required are orders of magnitude more massive than the mass of dust observed. 
We find that more massive extreme debris disks require proportionally larger colliding bodies. 
As a consequence, the most massive observed extreme debris disks require collisions of Mars- to Earth-mass bodies.
Extreme debris disks thus reveal the {ongoing formation of} rocky planets comparable in size to the rocky planets of our own solar system.

\end{abstract}


%%%%%%%%%%%%%%%%%%%%% INTRO %%%%%%%%%%%%%%%%%%%%%% 

\section{Introduction}%\label{s:intro}

\noindent Collisions between planetary bodies are fundamental to the growth of rocky planets. The final stage of terrestrial planet growth involves a few to tens of collisions between planetary embryos -- the so-called giant impact phase \citep{Chambers01,Kokubo10,Quintana16}. In our own solar system we have clear evidence of such collisions from their consequences: Earth's Moon \citep{Canup22}, the Pluto--Charon system \citep[][]{Canup05}; and many more pieces of evidence that strongly suggest giant impacts, e.g. Mercury's massive core \citep{Benz07,Chau18}, Borealis basin on Mars \citep[][]{Marinova08,Citron18}. Outside of our solar system there is now a growing body of evidence of recent giant impacts \citep[][]{Lisse09,Su19,Watt21,Moor22,Kenworthy23,Tzanidakis26}. 

Giant impacts are extremely energetic events \citep{Melosh90,Lock17,Carter20}. Yet directly observing a giant impact in progress is highly improbable; instead we rely on indirect signatures of such impacts. Observing a hot, post-impact body is possible (and has recently been achieved; \citealp{Kenworthy23}), but is also very unlikely due to their similarity to stellar temperatures and relatively short-lived nature. Here we deal instead with more readily detectable impact ejecta.

Material ejected from planetary collisions is unbound from any large planetary remnants, but generally remains gravitationally bound to the star. 
Giant impacts (a term which has no generally-accepted, specific definition) commonly involve a large degree of vaporization due to the high energies and velocities involved \citep[e.g.][]{Carter20}. However, even km-scale planetesimals can reach impact velocities sufficiently high to cause vaporization of silicates if influenced by other gravitating bodies \citep[][]{Davies20,CarterStewart20}. Material ejected into space that is in the vapor phase will rapidly expand and, due to the corresponding drop in pressure, condense into small particulate matter -- dust \citep{Benz07,Takasawa11,Johnson12}. It is this vapor condensate dust that has been observed in extreme debris disks \citep[][]{Lisse09,Meng14,Su19,Su25} and is likely responsible for the deep, long-duration eclipse seen in ASSASN-21qj \citep{Kenworthy23}.

Extreme debris disks are characterized by bright infrared excesses {(typical fractional luminosities $\gtrsim$0.01)} produced by warm dust emission %around other stars
\citep{Melis10,Lisse12,Moor21}. Some extreme debris disks show variability on year or shorter timescales indicating recent collisions \citep{Meng14,Su19,Rieke21,Moor22}. 
{The collisional cascade model for `classical' debris disks cannot explain this short term variability, nor sustain such high flux over the ages of extreme debris disk systems \citep{Wyatt07}. Instead, it is suggested that dust produced by high velocity, vaporizing collisions can explain the brightness and short timescale variability of extreme debris disks. An alternative scenario that has been proposed is the collisional avalanche model \citep{Grigorieva07}, in which the dust ejected from the destruction of asteroid-sized body induces a collisional chain reaction in an external disk; however, recent work has shown that this scenario is unlikely to produce the observed variations in luminosity \citep{Thebault18}.}
Much progress has been made in understanding how giant impact ejecta can produce the complex signals observed in extreme debris disks \citep{Johnson12Debris,Jackson14,Watt21,Lewis23,Watt24}; but there is as yet no clear way to relate the observed dust to the mass or size of the impacting bodies. Typically, the inferred mass of dust is totalled and converted to a body size assuming the implausible scenario that the entire mass of the body was turned into dust.

Here we carry out a suite of high-resolution hydrodynamic planetary collision simulations and obtain the mass 
and state of ejecta in order to relate extreme debris disks to the collisions that produced them. 
We develop a scaling relation that gives the mass of vaporized ejecta from collisions as a function of their kinetic energy. We show that the brightest extreme debris disks are the result of planet-scale collisions indicating planetary accretion and growth as opposed to disruption and erosion seen in classical debris disks.

%%%%%%%%%%%%%%%%%%%% METHODS %%%%%%%%%%%%%%%%%%%%%

\section{Numerical methods}\label{s:methods}

\subsection{Impact simulations}

We modelled collisions between strengthless planetary bodies primarily using the smoothed particle hydrodynamics (SPH) code SWIFT \citep{swift,Schaller24}. Simulations were carried out using the `planetary\_plus\_subtask\_speedup' version of SWIFT v0.9 \citep{swift-planetary_subtask} to take advantage of the speed improvements for planetary simulations. 
We additionally carried out a small subset of simulations with the planetary version of the SPH code Gadget2 \citep{Springel05,Carter22} for comparison. For this work the key differences between these two codes are the inclusion of a maximum smoothing length, or equivalently for equal-mass particle simulations a `density floor', in SWIFT; and while Gadget2-planetary uses a density-entropy formulation of SPH, SWIFT's planetary version uses a density-internal energy formulation. For the majority of the SWIFT simulations we set the maximum smoothing length such that the density floor (the minimum density a particle can have) is $\sim$10$^{-5}$\,g\,cm$^{-3}$. A subset of these simulations were repeated with lower density floors for comparison.

The majority of the simulations presented in this work have a numerical resolution of 10$^6$ particles per target mass. The masses of particles in both target and projectile bodies are set approximately equal, thus the number of particles used to represent the projectile depends on the projectile-to-target mass ratio. Subsets of simulations were repeated at higher resolutions of 5$\times$10$^6$ and 1$\times$10$^7$ for comparison. Remnant masses typically varied by less than 1\% between simulations with different numerical resolution, while vapor masses/fractions typically varied by less than 5\%.

All simulations in this work used two layer planets with a metallic core and rocky mantle with a core mass fraction of 0.3. We used the state-of-the-art ANEOS/M-ANEOS equations of state (EoS) from  \citet{StewartANEOSFo,StewartANEOSFe,StewartANEOSFeSi}. Planetary mantles are represented by forsterite \citep{Stewart19}, and planetary cores are represented by an iron-silicon alloy (85\% Fe, 15\% Si; \citealp{StewartANEOSFeSi}) for impacts involving a target of 0.01\mearth{} and above or pure iron \citep{StewartANEOSFe} for impacts with lower mass targets. The Fe-Si alloy was designed to provide a better match to the densities of Earth's core, which contains some lighter elements (likely including Si) in addition to the majority Fe and Ni. Since the partitioning of lighter elements into the metallic core is generally weaker at lower pressures \citep{Gessmann01} we chose to use pure iron for the smallest targets in our simulations. To avoid having any simulations in which there are two different core EoS, all projectiles use the same core EoS as their associated target (thus a body with a mass less than 0.01\mearth{} colliding with a body of mass e.g. 0.1\mearth{} uses the Fe-Si alloy core EoS despite the pure Fe likely being more appropriate for such a body). Nevertheless, we expect this simplification to have a negligible effect on our results.

We explore target masses ranging from 10$^{-3}$ to {1.5}\,\mearth{}, projectile-to-target mass ratios ($\gamma$) from 10$^{-3}$ to 1.0, impact parameters ($b$) from 0 to 0.966 (75\degrees{}), and impact velocities from 1.0 to {20} times the mutual escape velocity ($v_\mathrm{esc}$). The `core' simulation set contains every combination from target masses of 10$^{-3}$, 0.01, 0.1, and 1\,\mearth{}, $\gamma$ of 1$\times$10$^{-3}$, 0.01, 0.1, and 1, $b$ of 0.0, 0.26, 0.5, 0.707, and 0.87 (impact angles of 0\degrees{}, 15\degrees{}, 30\degrees{}, 45\degrees{}, and 60\degrees{}), and impact velocities of 1.0, 1.5, 2.0, 3.0, and 4.0 $v_\mathrm{esc}$ for equal-mass impacts for all but the smallest target, and 1.0, 3.0, 5.0, 7.5, and 10\,$v_\mathrm{esc}$ otherwise. The full list of {786} simulations is given in Table \ref{t:fullsimlist}. 

Planetary bodies were generated using the planetary impact toolkit ({planit}) package \citep{planit}. First an initial estimate of the central density is made (based on the required planetary mass), then a 1D planetary profile is built from the centre outward until the pressure drops to that specified for the surface, 10$^{-4}$\,GPa. If this planet does not have a mass sufficiently close to that required, the central density is adjusted 
and the process repeated until the required mass is reached (within a specified tolerance, 10$^{-3}$). Particles are placed according to these 1D profiles using SEAGen \citep{Kegerreis19}. 

All bodies were initialised with isentropic profiles for mantle and core. The mantle entropy was set to 3.03$\times$10$^7$\,J\,K$^{-1}$\,kg$^{-1}$, such that the entire mantle is solid with the surface just below the melt curve. A solid but hot mantle is likely the most appropriate initial state since the interval between collisions is expected to be much longer than the magma ocean freezing timescale \citep{Elkins-Tanton08}. The entropy of the target's core was set to a minimum of 1.81$\times$10$^7$\,J\,K$^{-1}$\,kg$^{-1}$ for the Fe-Si alloy and 1.80$\times$10$^7$\,J\,K$^{-1}$\,kg$^{-1}$ for pure iron cores in order to produce a fully liquid core appropriate for young terrestrial planets. However, if this entropy resulted in a target with a core that was colder than the mantle at the core-mantle boundary (a state we consider unrealistic), the entropy was increased (in steps of 0.01$\times$10$^7$\,J\,K$^{-1}$\,kg$^{-1}$) until the outer temperature of the core exceeded the temperature at the base of the mantle. In order to avoid the cores of smaller projectiles being in a different thermodynamic state and naturally buoyant, we set the core entropies of projectiles to match that of their associated target (thus projectiles can have cores colder than their mantles).

We also repeated a subset of the simulations with lower entropy starting conditions. For these simulations the standard mantle and core entropies were set to 2.70$\times$10$^7$\,J\,K$^{-1}$\,kg$^{-1}$ and 1.79$\times$10$^7$\,J\,K$^{-1}$\,kg$^{-1}$, respectively. These conditions are colder than typically used in similar SPH impact simulations (\citealp{Canup04,Cuk12,Lock17}) and should require greater energy input to cause vaporization. These lower entropy simulations generally lead to slightly lower ejecta vapor fractions, typically showing a difference less than 5\% from our nominal simulations. For collisions that produce very small ejecta and vapor fractions, the ejecta vapor fractions for the lower entropy simulations can be up to 10--20\% lower, however this difference in vaporized ejecta mass is comparable or smaller than the difference exhibited between collisions with similar energies (see section \ref{s:vaporizedejecta}).

The initialised bodies were equilibrated in isolation with a two-stage process. For the first 10\,hours the bodies had their initial entropies enforced and their particle velocities reduced by a factor of two each time step (using the `planetary\_veldamp' branch of SWIFT available from \citealp{swift-vel_damp}). The second stage allows the bodies to settle in isolation for a further 14\,hours. At the completion of the full 24\,hour equilibration process root-mean-squared particle velocities were less than 1\% of the body escape velocity in all cases.

Equilibrated planetary bodies were placed, using the corrected WoMa routines \citep{RuizBonilla21}, such that contact at the desired velocity and impact angle would occur one hour after the start of the collision simulation (if no tidal deformation occurred). Simulations were run for a total duration of 48\,hours, the end state thus corresponds to approximately 47\,hours after the collision began. {The SWIFT simulations use a cubic box, which was set to have side lengths of 10,000\,$\mathrm{R}_\oplus$, with the centre of mass of the impactors at the origin. Particles passing beyond the boundary of the box are removed from the simulation. In cases where a significant number of particles left the box, the simulation was restarted with a larger box. We incude in our analysis only simulations in which less than one percent of the mass was removed in this way.} The standard 10$^6$ particle resolution simulations took between 3\,hours and 76\,days (median $\sim$11\,days) to complete on 16 cores on the University of Bristol's BlueCrystal and BluePebble supercomputers.

\subsection{Analysis}\label{s:analysis}

The ten most massive remnants at the end of each simulation were identified by iteratively comparing particle kinetic energies to the gravitational potential energy of the collection of particles with the deepest potential well (see \citealp{Carter18} and \citealp{Dou24a} for more details). Each time the mass of the remnant converged to within 0.1\% those particles were assigned the next available remnant number and the calculation repeated with those particles ignored. The gravitational potential used to identify seeds for remnants was recalculated each time a remnant was found. Identified bound remnants with fewer than 500 particles were treated as part of the unbound ejecta.

Since we are interested primarily in the mass of dust that is assumed to rapidly condense from vaporized ejecta, we calculate vapor fractions assuming release to low pressure rather than using the instantaneous pressure of the particle in a particular snapshot (Figure \ref{f:phasedecompress}). The vapor fraction of each particle is calculated from its entropy, at that instant in time, assuming that the parcel of mass the particle represents decompresses isentropically to a release pressure of 5.2\,Pa (the triple point of forsterite; \citealp{Nagahara94}). For particles with entropies between that of incipient vaporization and complete vaporization at this release pressure, i.e. those inside the vapor dome, we calculate the vapor fraction using the lever rule. Particles with entropies below incipient melting or above complete melting are assigned vapor fractions of zero and one respectively. The total vapor fraction for a bound remnant or unbound ejecta is the mass-weighted sum of the individual particle vapor fractions. {Generally we use the vaporized ejecta mass calculated at the end of each simulation (48\,hours).} Remnant and vapor fraction calculations are implemented in the {planit} package \citep{planit}. 

The total vapor fraction of ejecta (or a bound remnant) can therefore vary both due to changes in the entropy of any individual constituent particle or changes in which particles are identified as bound/unbound. While, in general, these quantities should (and do) converge, when we are considering small numbers of particles in particular there can be variation in the calculated values between output snapshots. We therefore calculated the ejecta (and remnant) mass and vapor fraction across the final 10\,hours of simulation time and took the mean and standard deviation of these quantities as the final values and a measurement of the variation in those values. We treat these standard deviations similarly to an uncertainty though it should be noted that they are not `true' uncertainties and do not quantify the full uncertainty that would be contributed by uncertainties in factors such as EoS input parameters, impact parameters due to tidal deformation, density floor, artificial viscosity, and neighbour count. In most cases these calculated `uncertainties' are negligible and too small to be visible in figures.

Since the uncertainties on the data are so small, in order to calculate a measure of uncertainty or variation in the fit for the derived vapor mass scaling relation we use the bootstrap resampling method. We resample the vapor mass dataset 10,000 times (with replacement) and fit each of these 10,000 datasets independently. The final scaling relation parameters and uncertainties (Table \ref{t:fitparams}) are calculated as the mean and standard deviation of the 10,000 instances of the fit parameters.

%%%%%%%%%%%%%%%%%%%% RESULTS %%%%%%%%%%%%%%%%%%%%%

\section{Results}

\subsection{Impact evolution}

\begin{figure*}%[!hbt]
\centering\includegraphics[width=\textwidth]{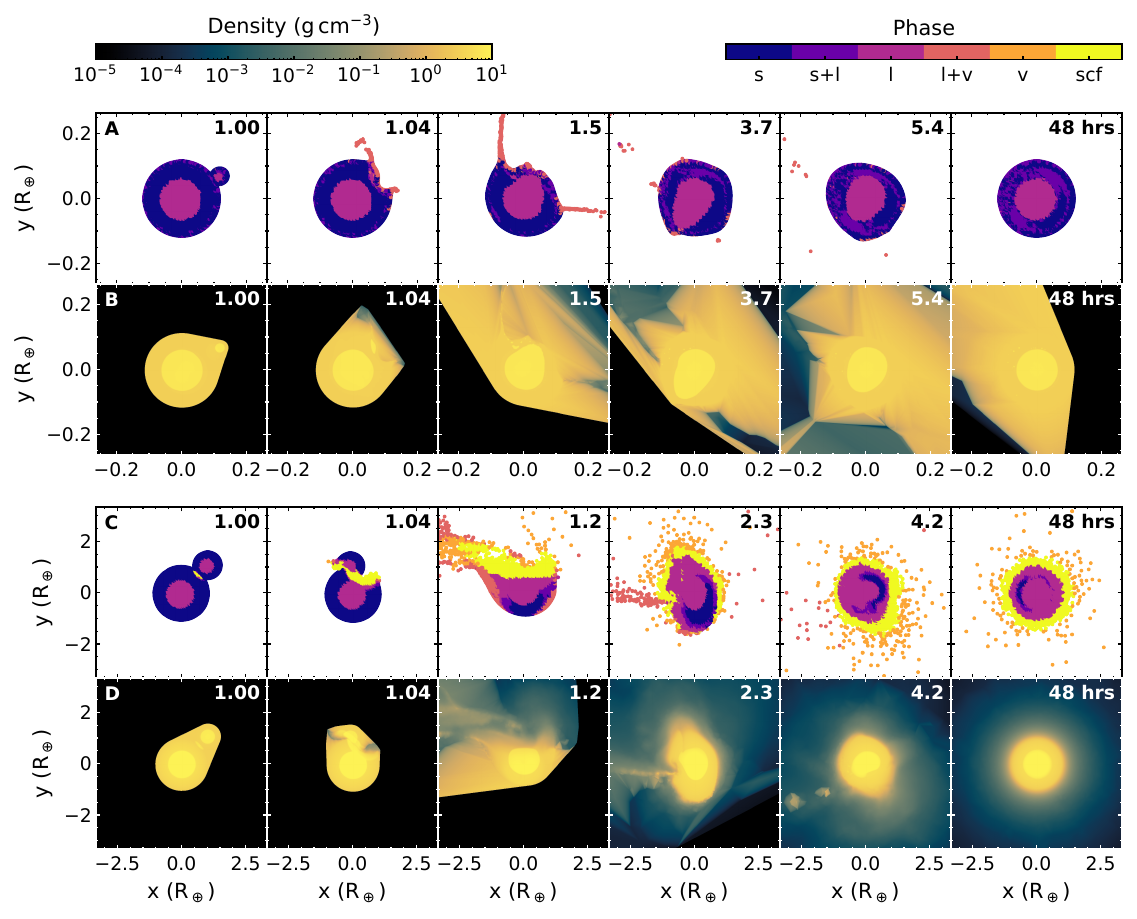}%0.94

\caption{{\bf Progression of two example simulated collisions.}
Instantaneous phase (solid [s], mixed solid and liquid [s+l], liquid [l], mixed liquid and vapor [l+v], vapor [v], or super critical fluid [scf]) and density in the midplane as two example collisions progress.  The number in the upper right of each panel is the simulation time in hours. Phase panels show SPH particle positions, where white indicates no particles exist at that location. Density panels show the fluid representation using (3D) interpolation, and so density can be defined between particle positions; black areas are either at or below the minimum shown on the color scale, or indicate regions in which there are no particles that allow the density to be interpolated. 
Rows A, B: an accretion collision with a 10$^{-3}$\,\mearth{} target and a mass ratio of 10$^{-2}$ at 3\,$v_\mathrm{esc}$ and an impact angle of 30$^\circ$ ($b = 0.5$). %cM1e-3p0mr1e-2v3.00b0.500N1e6Fo70S303Fe30S193. 
Rows C, D: a mildly erosive collision with a 1\,\mearth{} target and a mass ratio of 0.1 at 5\,$v_\mathrm{esc}$ and an impact angle of 45$^\circ$ ($b = 0.707$). %cM1p0mr1e-1v5.00b0.707N1e6Fo70S303FeSi30S181. 
Note a different scale is used for each collision. Corresponding materials plots are shown in Figure \ref{f:collseqmat}. \label{f:collseq}}
\end{figure*}

Our collision simulations span a large range of outcomes from near-perfect merging to highly erosive \citep{Leinhardt12}. Collisions with small projectile-to-target mass ratios (rows A and B in Figure \ref{f:collseq}) cause only moderate disturbance of the target and eject little mass (up to a few percent of the combined mass) unless the impact velocities are very high. Yet even the lowest energy collisions in our sample result in some melting and vaporization near the impact site (orange-red particles in Figure \ref{f:collseq}A, second frame onward).
Hit-and-run collisions result in very limited exchange of mass or accretion (Figure \ref{f:collseqmat}C and rows A and B in Figure \ref{f:collseq2}). However, hit-and-run collisions can cause substantial melting and vaporization at the impact site, in some cases resulting in a large degree of mantle melting with a liquid (magma ocean) surface layer and a low-mass, low-density vapor atmosphere (Figure \ref{f:collseq2}A).

Higher energy collisions often lead to erosion of the target body, and cause much more thermal alteration with a large amount of melting and vaporization 
(second collision, rows C and D, in Figure \ref{f:collseq}). The targets also suffer substantial `opening-up' and decompression, leading to large `streams' of reaccreting material later in the evolution ($>$2\,hours). These streams can cause significant further heating and vaporization as they fall onto the remnant body \citep{Carter20}. The resulting planets are typically very hot and inflated compared to the pre-impact state, with partially vaporized atmospheres of rock or even iron (Figure \ref{f:collseqmat}D and Figure \ref{f:collseq2}C).

The largest post-collision remnant body has mostly gravitationally settled by 48\,hours in all collisions (final column in Figure \ref{f:collseq}), including those that are highly disruptive. The mass of the largest remnant, the mass of the ejecta, and related properties typically converge about 20\,hours after the impact began.

\subsection{Vaporized ejecta}\label{s:vaporizedejecta}

Across orders of magnitude in both sizes of colliding bodies and impact energy, all simulated collisions produce some ejecta.  
The phases of particles extracted directly from the simulations are the `instantaneous' phase -- the properties of the material at the pressure it was experiencing at that moment in time (shown in Figure \ref{f:collseq}). Any unbound ejecta from a collision is expected to continue to expand into empty space causing its pressure to drop.  
We assume that decompression of ejecta occurs rapidly compared to the cooling timescale and thus calculate the vapor fraction of ejecta assuming isentropic decompression to the triple point (5.2\,Pa; Figure \ref{f:phasedecompress}).
{The total vaporized ejecta mass is the sum of the individual particle vapor masses (vapor fraction multiplied by particle mass). The vapor mass/fraction is calculated at the end time of the simulation except where an earlier time is directly specified.}
For all but the slowest oblique collisions involving the smallest targets some fraction of the ejecta is vaporised. At the highest impact energies large fractions of the total colliding mass are vaporised and ejected (Figure \ref{f:Q_ejvapfrac}).

The vapor fraction of the ejecta shows no clear correlation with specific impact energy (standard, normalised or modified) or any of the impact parameters. The only clear trends are that very high vapor fractions do not occur at the lowest impact energies, and low vapor fractions do not occur at the highest energies.

\begin{figure}
\centering\includegraphics[width=\columnwidth]{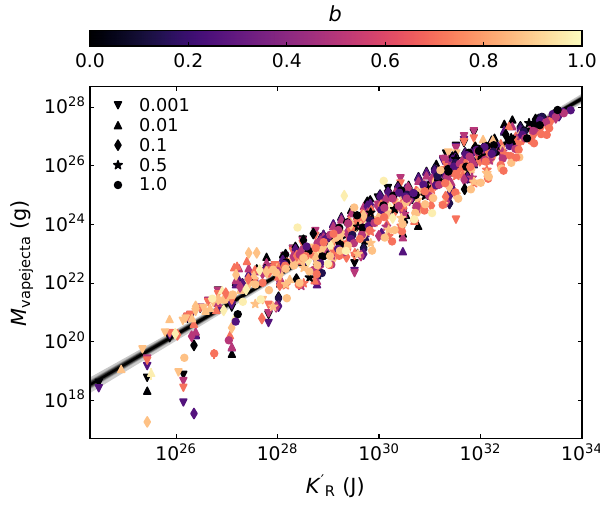}
\caption{{\bf The mass of vaporized ejecta scales with the modified kinetic energy of the collision.}
Mass of vaporized ejecta, $M_\mathrm{vapejecta}$, as a function of modified kinetic energy, $K'_\mathrm{R}$. The shape of symbols corresponds to the mass ratio of the collision and the colour corresponds to the impact parameter, $b$. The solid black line and grey band indicate the best fitting power law and the range of fits obtained via bootstrap fitting. The amplitude is $1.22^{+1.39}_{-0.65}\times 10^{-6}$ and the power law index is 1.01$\pm$0.01. The full fit parameters are provided in Table \ref{t:fitparams}. Note the `uncertainties' (derived from the final 10\,hours of simulation time, see section \ref{s:analysis}) are too small to be seen for the majority of collisions. \label{f:Mvap_K}}
\end{figure}

We find that the mass of vaporized ejecta increases with increasing kinetic energy of the collision. We define a `modified kinetic energy', based on \citep{Leinhardt12}, that accounts for the energy provided by the geometrically overlapping portion of the projectile in oblique and unequal-mass collisions,
\begin{equation}
    K'_\mathrm{R} = Q'_\mathrm{R} M_\mathrm{tot} = \frac{1}{2} \mu_\alpha {v^2} = \frac{1}{2} \frac{\alpha \gamma}{(\alpha \gamma +1)} M_\mathrm{targ}  {v^2},
    \label{e:K_def}
\end{equation} 
where $Q'_\mathrm{R}$ is the modified specific impact energy
(see appendix \ref{a:Qp}), $M_\mathrm{tot}$ is the total mass, $\mu_\alpha$ is a modified reduced mass that accounts for the interacting portion of the projectile's mass, $\alpha$ (see \citealp{Leinhardt12} and appendix \ref{a:alpha}), $v$ is the impact velocity, and $M_\mathrm{targ}$ is the mass of the target body. 
The absolute mass of vaporized ejecta scales close to linearly with this modified kinetic energy (Figure \ref{f:Mvap_K}) and is well fit by a power law:
\begin{equation}
    \left(\frac{M_\mathrm{vapejecta}}{\mathrm{g}}\right) = 1.22\times 10^{-6} \, \left(\frac{K'_\mathrm{R}}{\mathrm{J}}\right)^{1.01}.
    \label{e:vapejecta_K}
\end{equation} 
We use a bootstrap fitting procedure to determine the parameters. While there is a range of vaporized ejecta mass at any given energy, the range of power law fits obtained from the bootstrap fitting is narrow (Table \ref{t:fitparams}). 
The mass of vaporized ejecta produced by any collision can thus be predicted from its basic collision parameters.
%

%%%%%%%%%%%%%%%%%%% DISCUSSION %%%%%%%%%%%%%%%%%%%%

\section{Discussion}\label{s:discussion}

The vaporized ejecta masses measured from the simulations are well converged, with little variation throughout the final 10\,hours of simulation time (most of the `error bars' in Figure \ref{f:Mvap_K} are too small to be visible). The vapor masses or fractions obtained from simulations conducted at different numerical resolutions, or with lower initial entropies, or with a different hydrodynamic code, show good agreement, generally within a few percent (see section \ref{s:methods}), though substantial differences sometimes occur for the smallest ejecta masses. Taken together, we can thus be confident in the accuracy of the vapor masses and the vaporized ejecta mass scaling relation (Figure \ref{f:Mvap_K} and equation \ref{e:vapejecta_K}).

Since we have used a very refractory silicate, forsterite, to model the mantles of colliding bodies, the measured vapor masses are likely to be underestimated compared to real planetary mantles which would contain a variety of minerals. For the smallest bodies, which may not have undergone complete differentiation, any porosity would likely also make vaporization easier and increase the vapor mass. However, we have also neglected material strength which may somewhat reduce the mass of material vaporized or the mass of ejecta. Material strength would be more significant for the lowest mass bodies, collisions with which generally produce little vaporized ejecta. Since our target bodies are all well within the gravity-dominated regime, we consider the effect of neglecting material strength to be minor.

In the simulations that produce the smallest ejecta masses ($\lesssim$10$^{-5} \, M_\mathrm{tot}$) the total number of particles representing the ejecta is low ($\sim$10 or fewer) which would generally be considered unresolved. Since these make up only a small fraction of the simulations, they appear to match well with the trend for the rest of the simulations (Figure \ref{f:Q_Mejecta}), and low vaporized ejecta masses have little effect on our results, we include these low ejecta mass results despite the low particle count. 

It is evident in Figure \ref{f:phasedecompress} that many ejecta particles have reached the density floor of our simulations (these particles lie on low pressure curves in pressure–entropy space). Since these are ejected (unbound) particles that are expected to travel away from the remnant planetary bodies and have few further interactions, these particles reaching the minimum density and becoming essentially ballistic objects should not cause thermodynamic problems.  We recalculate the phase and vapor fractions of ejecta particles assuming decompression to the triple point so the density floor should not affect our measured vaporized ejecta masses.

The vaporised ejecta mass scaling relation (equation \ref{e:vapejecta_K}) has a power law index very close to unity (1.01$\pm$0.01). Since we do not expect, a priori, a linear relation between the modified kinetic energy and the mass of vaporised ejecta we choose to retain the power law index (gradient) as a free parameter rather than fixing it to one.

\subsection{Linking dust mass to colliding bodies}

\begin{figure*}
\centering\includegraphics[width=0.94\textwidth]{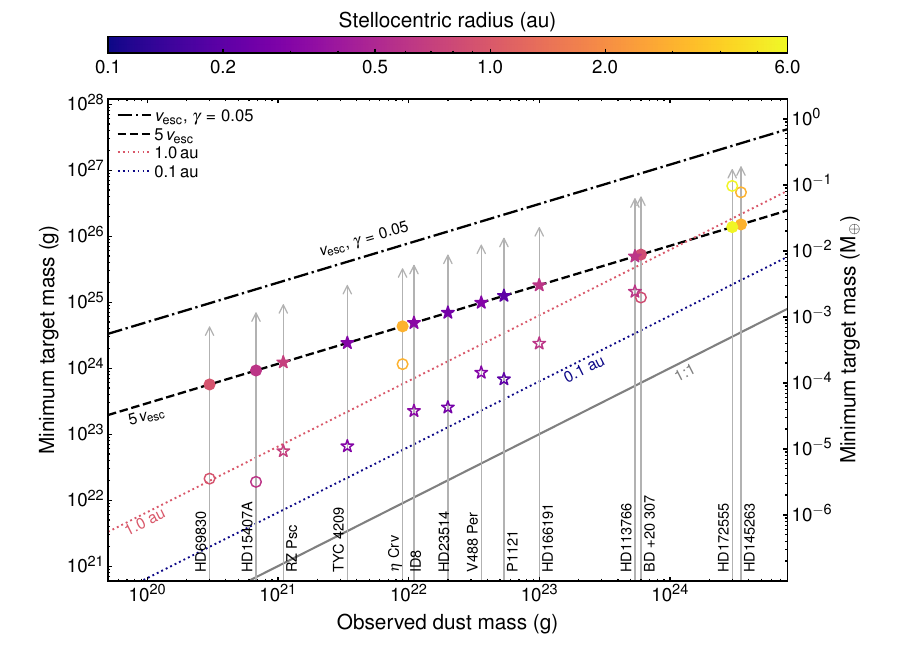}
\caption{
{\bf The observed extreme debris disks require planet-mass collisions.}
Minimum target mass needed to produce observed dust mass. Black lines show scenarios for the collision velocity being a multiple of the mutual escape velocity, coloured lines show scenarios with maximally eccentric colliders at different distances from their host star. Lines are for collisions at 45$^\circ$, with a mass ratio of 0.5, around a one solar mass star except where indicated otherwise. Symbols show the minimum target mass calculated from inferred dust masses observed in extreme debris disks for a velocity five times the escape velocity (filled) and the maximally eccentric scenario (open). These points assume collisions at 45$^\circ$ with a mass ratio of 0.5. Symbols are also coloured according to the observationally derived minimum stellocentric radius of the emission. Stars indicate those systems that show significant variability.\label{f:EDD_Mtarg}}
\end{figure*}

The modified kinetic energy depends on the target mass, the impact velocity, the mass ratio, and the interacting mass -- which itself depends on the mass ratio and the impact parameter. Since the mass of vaporized ejecta scales with modified kinetic energy, the mass of vaporized ejecta therefore depends on the same quantities. For a given collision scenario -- impact velocity, angle, and mass ratio -- there is therefore a direct relation between the mass of the target and the mass of vaporized ejecta which will be observable as dust.

While we cannot know exactly the parameters of a collision that occurred around another star, if we assume a maximum impact velocity and reasonable impact angle and mass ratio we can derive a minimum target mass required to produce a given mass of dust, $M_\mathrm{dust}$. We now consider two scenarios, one using a multiple of the escape velocity for which we choose a maximum typically expected velocity, and one that provides essentially the most extreme impact velocity possible for two co-planar orbiting bodies. 

For the first scenario the impact velocity is taken as some multiple of the mutual escape velocity, $v_\mathrm{esc}$. The minimum target mass %(in Earth masses, M$_\oplus$) 
is then defined,
\begin{multline}
    \left(\frac{M_\mathrm{targ}}{\mathrm{M_\oplus}}\right) 
    = 8.28 \times 10^{-5}\,  \left(\frac{M_\mathrm{dust}}{10^{20}\,\mathrm{g}}\right)^{0.596} 
    \left(\frac{v}{v_\mathrm{esc}}\right)^{-\frac{6}{5}} \\
    \left(\frac{\left(1+\gamma^\frac{1}{3}\right)}{ \left( 1+\gamma \right) } \right)^\frac{3}{5} 
    \left( \frac{(\alpha \gamma +1)}{\alpha \gamma} \right)^{\frac{3}{5}} 
    \left(\frac{\rho}{5\,\mathrm{g\,cm^{-3}}}\right)^{-\frac{1}{5}},
\end{multline}
where $\rho$ is the average density of the target and projectile (assumed to be equal). We choose a nominal impact angle of 45$^\circ$, mass ratio of 0.5, distance from the star of 1\,au, and an impact velocity of five times the mutual escape velocity. 
This scenario using a multiple of $v_\mathrm{esc}$ is illustrated with the dashed and dash-dotted lines in Figure \ref{f:EDD_Mtarg}.

\begin{table*}
\caption{{\bf Properties of known extreme debris disk systems and calculated minimum collider masses.} The published stellar {mass, approximate age of the star}, minimum disk radius, and minimum observed dust mass for known extreme debris disks with measured dust masses. Also shown are the minimum target mass required to explain the dust mass calculated from our two collision scenario models combined with our derived vaporized ejecta mass scaling relation. \label{t:EDD}}
%\footnotesize
\vspace{-4mm}
\begin{center}
\begin{tabular}{lrrrrcrrr}
\hline %\toprule
System  & Stellar mass, & {Age} & Inner & Observed mass, & Variable & \multicolumn{2}{c}{Minimum target mass (\mearth{})} & Refs \\
 &   $M_*$ (M$_\odot$) & {(Myr)} &  radius (au)  & $M_\mathrm{dust}$ (g)  & & at 5\,$v_\mathrm{esc}$ & maximally eccentric &  \\
\hline %\midrule
HD69830  &  0.86  &  7500  &  0.93  &  3$\times$10$^{20 }$ &    &  9.49$\times$10$^{-5 }$ &  3.55$\times$10$^{-6 }$ &  \postnote{\citealp{Lisse07}},\postnote{\citealp{Tanner15}} \\
HD15407A  &  1.4  &  80  &  0.6  &  6.8$\times$10$^{20 }$ &    &  1.55$\times$10$^{-4 }$ &  3.17$\times$10$^{-6 }$ &  \postnote{\citealp{Fujiwara12}} \\
RZ Psc  &  0.9  &  30  &  0.7  &  1.1$\times$10$^{21 }$ &  Y  &  2.06$\times$10$^{-4 }$ &  9.29$\times$10$^{-6 }$ &  \postnote{\citealp{Su23}},\postnote{\citealp{deWit13}} \\
TYC 4209  &  1.0  &  275  &  0.3  &  3.4$\times$10$^{21 }$ &  Y  &  4.04$\times$10$^{-4 }$ &  1.10$\times$10$^{-5 }$ &  \postnote{\citealp{Moor22}} \\
$\eta$ Crv  &  1.5  &  1400  &  3.0  &  9$\times$10$^{21 }$ &    &  7.21$\times$10$^{-4 }$ &  1.93$\times$10$^{-4 }$ &  \postnote{\citealp{Lisse12}} \\
ID8  &  1.0  &  35  &  0.32  &  1.1$\times$10$^{22 }$ &  Y  &  8.13$\times$10$^{-4 }$ &  3.77$\times$10$^{-5 }$ &  \postnote{\citealp{Meng14}} \\
HD23514  &  1.25  &  100  &  0.25  &  2$\times$10$^{22 }$ &  Y  &  1.16$\times$10$^{-3 }$ &  4.26$\times$10$^{-5 }$ &  \postnote{\citealp{Rhee08}} \\
V488 Per  &  0.8  &  80  &  0.3  &  3.6$\times$10$^{22 }$ &  Y  &  1.65$\times$10$^{-3 }$ &  1.43$\times$10$^{-4 }$ &  \postnote{\citealp{Rieke21}} \\
P1121  &  1.0  &  80  &  0.2  &  5.37$\times$10$^{22 }$ &  Y  &  2.09$\times$10$^{-3 }$ &  1.14$\times$10$^{-4 }$ &  \postnote{\citealp{Su19}} \\
HD166191  &  1.6  &  10  &  0.6  &  1$\times$10$^{23 }$ &  Y  &  3.03$\times$10$^{-3 }$ &  3.96$\times$10$^{-4 }$ &  \postnote{\citealp{Su22}} \\
HD113766  &  1.4  &  17  &  0.6  &  5.4$\times$10$^{23 }$ &  Y  &  8.28$\times$10$^{-3 }$ &  2.42$\times$10$^{-3 }$ &  \postnote{\citealp{Olofsson13}},\postnote{\citealp{Su20}} \\
BD +20 307  &  2.7  &  1000  &  0.85  &  6$\times$10$^{23 }$ &    &  8.82$\times$10$^{-3 }$ &  1.97$\times$10$^{-3 }$ &  \postnote{\citealp{Weinberger11}} \\
HD172555  &  1.86  &  20  &  5.8  &  3$\times$10$^{24 }$ &    &  2.30$\times$10$^{-2 }$ &  9.67$\times$10$^{-2 }$ &  \postnote{\citealp{Su20}},\postnote{\citealp{Lisse09}} \\
HD145263  &  1.4  &  11  &  3.0  &  3.5$\times$10$^{24 }$ &    &  2.53$\times$10$^{-2 }$ &  7.75$\times$10$^{-2 }$ &  \postnote{\citealp{Lisse20}} \\
\hline %\bottomrule
\end{tabular}
\end{center}
\printpostnotes
\end{table*}
For the second collision scenario we choose an extreme impact in which a maximally eccentric body collides with a body on a circular orbit orbiting in the same sense and plane. The impact velocity is thus the pericenter, $q$, velocity (equal to the escape velocity from the star) minus the Keplerian velocity at the radius of the collision, $a = q$:
\begin{equation}
    v = \sqrt{\frac{2GM_*}{q}} - \sqrt{\frac{GM_*}{a}} = \left(\sqrt{2}-1 \right) \sqrt{\frac{GM_*}{a}},
\end{equation}
where $G$ is the gravitational constant and $M_*$ is the stellar mass. We ignore the additional velocity acquired due to the gravitational attraction between the colliding bodies as it will generally be negligible compared to the orbital velocity. 
Combining the above with our vapor mass scaling relation (equation \ref{e:vapejecta_K}), we find the minimum target mass, 
\begin{multline}
%\begin{equation}
    \left(\frac{M_\mathrm{targ}}{\mathrm{M_\oplus}}\right)  = 1.25 \times 10^{-7}\, \left(\frac{M_\mathrm{dust}}{10^{20}\,\mathrm{g}} \right)^{0.99} \\
    \left(\frac{a}{\mathrm{au}} \right) \left( \frac{\mathrm{M}_\odot}{M_*} \right) \frac{(\alpha \gamma + 1)}{{\alpha \gamma}}.
%\end{equation}
\end{multline}
%
%\nopagebreak
This extreme collision scenario is illustrated with the coloured dotted lines in Figure \ref{f:EDD_Mtarg}. 
It should be noted that this is a highly improbable collision and more likely collisions would have substantially lower impact velocities, and hence require much higher target masses to produce the same dust mass.

{\subsection{Non-vaporized ejecta}

In this work we have focussed on the vaporized portion of the ejecta and ignored the non-vaporized portion. The vaporized ejecta will produce the first and fastest-evolving observable signature, and for the highest energy impacts also the brightest infrared excess. However, the non-vaporized ejecta may also produce observable dust. 

\citet{Watt24} explored the evolution of $\sim$50\,km and larger sized fragments resulting from giant impacts. They calculated the mass of these solid fragments from all ejected particles with a vapor fraction of 0.1 or lower in giant impacts involving approximately Mars-mass planets. They found that this fragment population can be collisionally active and produce observable dust over longer timescales than the vaporized ejecta similarly to a traditional collisional cascade. \citet{Watt24} also found that the timing, lifetime, and brightness of this `sustained' extreme debris disk is strongly influenced by the distance from the star at which the giant impact occurred.

The remaining portion of the debris, the melted material from particles with vapor fractions less than 1, has a complex evolution. As a mixed phase (liquid and vapor) parcel of material decompresses into empty space the vapor portion will expand in volume rapidly by many orders of magnitude \citep{Davies19b,Davies20}, and even for small vapor fractions the volume is dominated by the vapor. For lower entropies, the parcel of mass is initially dominated by melt as it decompresses through the vapor dome and the evolution is dominated by the formation and expansion of bubbles of vapor. These rapidly expanding bubbles likely act to break up any melt into small droplets \citep{Melosh91}. For high entropies, the parcel is dominated by vapor as it enters the vapor dome and the melt forms as droplets condensing from the vapor \citep{Johnson12,Davies19}. 

In both cases we have melt droplets in an orders-of-magnitude larger volume of vapor. As the system continues to decompress and cool the melt freezes and the vapor condenses into dust. The sizes of melt droplets and vapor condensate depend on the impact velocity and the size of the melt/vapor plume \citep{Johnson12}. It is unclear at what vapor fraction the vapor no longer dominates the evolution, the 10\% value used by \citet{Watt24} is an approximation. The `melt' portion of ejecta may thus produce one or more further narrow size distribution `dust' populations that can contribute to the infrared flux observed from extreme debris disk and may play an important role in the complex variability seen in some disks \citep[e.g.][]{Su19}. The relative proportions of ejecta that form large fragments vs melt droplets vs vapor is highly dependent on the impact (see Figure \ref{f:Q_ejvapfrac_noncorr}), but in all cases it is the vapor condensate that would be expected to be seen first.}

\subsection{Extreme debris disks require colliding planets}

We find that the minimum masses of the colliding planets required to produce the dust measured in extreme debris disks are all orders of magnitude greater than the observed dust mass for both collision scenarios (Figure \ref{f:EDD_Mtarg}, colored symbols). 
Extreme debris disks therefore require colliding bodies much more massive than those inferred using the typical assumption that the observed dust mass is equal to the mass of the impacted body (solid grey '1:1' line in Figure \ref{f:EDD_Mtarg}). 
Rather than 10--100\,km scale asteroids \citep{Lisse12,Meng14,Su19,Rieke21}, Ceres- to Moon-sized (1000--3500\,km diameter) or larger colliders are needed to explain the observations. For the two most massive extreme debris disks, HD172555 and HD145263, collisions with at least Mars-mass and more likely Earth-mass planets are required to produce the observed disk in a single collision. Thus, unlike typical debris disks in which we see the remains from planet formation being ground down, in extreme debris disks we are {typically} observing the results of active planet formation.

{Extreme debris disks are usually found around young stars (up to a few hundred million years old), when we might expect that planet formation is still ongoing. Indeed, the ages of the majority of the stars in the sample discussed here are less than 200\,Myr (see Table \ref{t:EDD}). The most massive disks and all those with significant observed variability are young, consistent with the `extreme' properties of these disks being due to giant impacts during the planet formation phase. Three of the systems, however, are much older ($\gtrsim$1\,Gyr), and would not be expected to still be undergoing planet formation. A more likely scenario for these systems is a `late instability' \citep[e.g.][]{Smith09,Kaib16} causing a new period of major collisional activity.}

Typical giant impacts in $N$-body simulations of planet formation have velocities 1--4\,$v_\mathrm{esc}$ \citep{Kokubo10,Quintana16}. We choose 5\,$v_\mathrm{esc}$ as a high, but reasonable impact velocity for most systems that provides a conservative estimate for the minimum target mass. We expect the real masses of the colliding bodies that produce extreme debris disks to be greater than calculated assuming a collision at 5\,$v_\mathrm{esc}$ (i.e. between the dashed and dot-dashed lines in Figure \ref{f:EDD_Mtarg}). For disks around Sun-like stars with inner edges at $\sim$1\,au the minium target masses calculated for the two scenarios cross at a dust mass of $\sim1.5 \times 10^{24}$\,g. Above this dust mass, the impact velocity in the extreme eccentric collision scenario drops below 5\,$v_\mathrm{esc}$ for the required mass and should be used to obtain the minimum mass, as is the case for the two most massive disks (Figure \ref{f:EDD_Mtarg}). 

The spectra of extreme debris disks also provide important clues about their origins, and act as a diagnostic to confirm the vaporization demonstrated in our simulations. Several disks show refractory dominated compositions with SiO$_2$-rich species indicative of high-temperature processing (HD172555, \citealp{Lisse09}; $\eta$~Corvi, \citealp{Lisse12}; HD15407A, \citealp{Fujiwara12}; HD23514, \citealp{Su25}), as would be expected to occur via vaporizing collisions. HD172555 (one of the most massive known extreme debris disks) also shows evidence of SiO vapor in its spectrum \citep{Lisse09,Johnson12Debris}. The presence of SiO provides further strong evidence for a vaporizing planetary collision. Future work should examine the thermodynamic pathways of the ejecta from the impacts modelled here to constrain the expected composition and mineralogy of the observable dust.

Not all extreme debris disks have shown significant variability during the timescales over which they have been observed (up to decades), and those that do show variability have not faded to zero excess infrared emission \citep{Su19,Moor22}. The observed dust in an extreme debris disk may therefore not be caused by a single (giant) impact, and there may be additional dust produced over longer timescales due to condensed portions of ejecta \citep{Watt24}. In all cases we have used the minimum dust mass inferred from the observations (and the minimum disk radius), the actual dust mass in these systems may be substantially larger, so even allowing for multiple collisions our minimum target mass estimates are still reasonable. In the case of RZ Psc the dust mass we use is an estimate of the change in mass over $\sim$year-long timescales \citep{Su23} rather than the much larger total mass of dust in the disk; therefore, a single collision involving a body with a minimum mass greater than that of Ceres is required. 

While the derived minimum masses are not unique, varying the mass ratio and impact angle/parameter will cause some shift, using the most common impact angle (45$^\circ$) and a high mass ratio (0.5) make our nominal scenarios fairly extreme, conservative, but plausible collisions. We therefore expect that in the majority of observed extreme debris disk systems the actual masses of the colliding bodies would be significantly higher than our calculated minima. Thus, extreme debris disks evidence giant impacts between solar system like rocky planets in forming exoplanet systems. Since large planetary bodies are required to rapidly produce fairly modest masses of observable dust and the lifetime of small dust is short, the low frequency of very massive debris disks does not mean the frequency of giant impacts is necessarily also low, and giant impacts between rocky planets may be more common than previously estimated \citep{Jackson12,Genda15DebrisDisks}.

%\clearpage
%\newpage

\begin{acknowledgments}
We thank the anonymous reviewer for helpful comments that improved this manuscript. We thank G. Collins, T. Elliott, M. Lodge, P. Skrzypczyk, and M. Kenworthy for helpful discussions. 
This research has made use of NASA's Astrophysics Data System. 
This work was carried out using the computational facilities of the Advanced Computing Research Centre, University of Bristol -- \url{http://www.bristol.ac.uk/acrc/}.

PJC and ZML acknowledge financial support from the UK Science and
Technology Facilities Council (grant numbers: ST/V000454/1 and
ST/Y002024/1). This research was supported in part by grant no. NSF PHY-2309135 to the Kavli Institute for Theoretical Physics (KITP). 

Simulation initial conditions, final states, and processed data used in this work are available on the \textit{Harvard Dataverse}, (doi.org/doi:10.7910/DVN/PFSTEF) \citep{Carter26Data}. 
The full datasets generated in this work are available from the corresponding author on reasonable request.

Codes necessary to reproduce the analysis presented in this article are available on \textit{GitHub}, (github.com/PhilJCarter/ImpactVaporEjecta) \citep{Carter26Exec}.

\end{acknowledgments}

\begin{contribution}

PJC and ZML came up with the initial research concept. PJC ran the simulations and carried out the analysis. PJC was responsible for drafting and submitting the manuscript. PJC and ZML edited the manuscript.

\end{contribution}

\software{SWIFT \citep{swift,swift-vel_damp,swift-planetary_subtask}, 
    Gadget2 \citep{Springel05,Carter22}, 
    planit \citep{planit}, 
    SEAGen \citep{Kegerreis19}, 
    WoMa \citep{RuizBonilla21}, 
    NumPy \citep{numpy}, 
    SciPy \citep{scipy}, 
    matplotlib \citep{matplotlib}, 
    Jupyter \citep{jupyter}
          }

%\bibliography{references}
%\bibliographystyle{aasjournal}

\appendix

\setcounter{figure}{0}
\setcounter{table}{0}
\renewcommand{\thefigure}{A\arabic{figure}}
\renewcommand{\thetable}{A\arabic{table}}
\renewcommand{\theHtable}{A.\thetable}
\renewcommand{\theHfigure}{A.\thefigure}

\section{Example impacts}

\begin{figure*}%[!hbt]
\centering\includegraphics[width=0.94\textwidth]{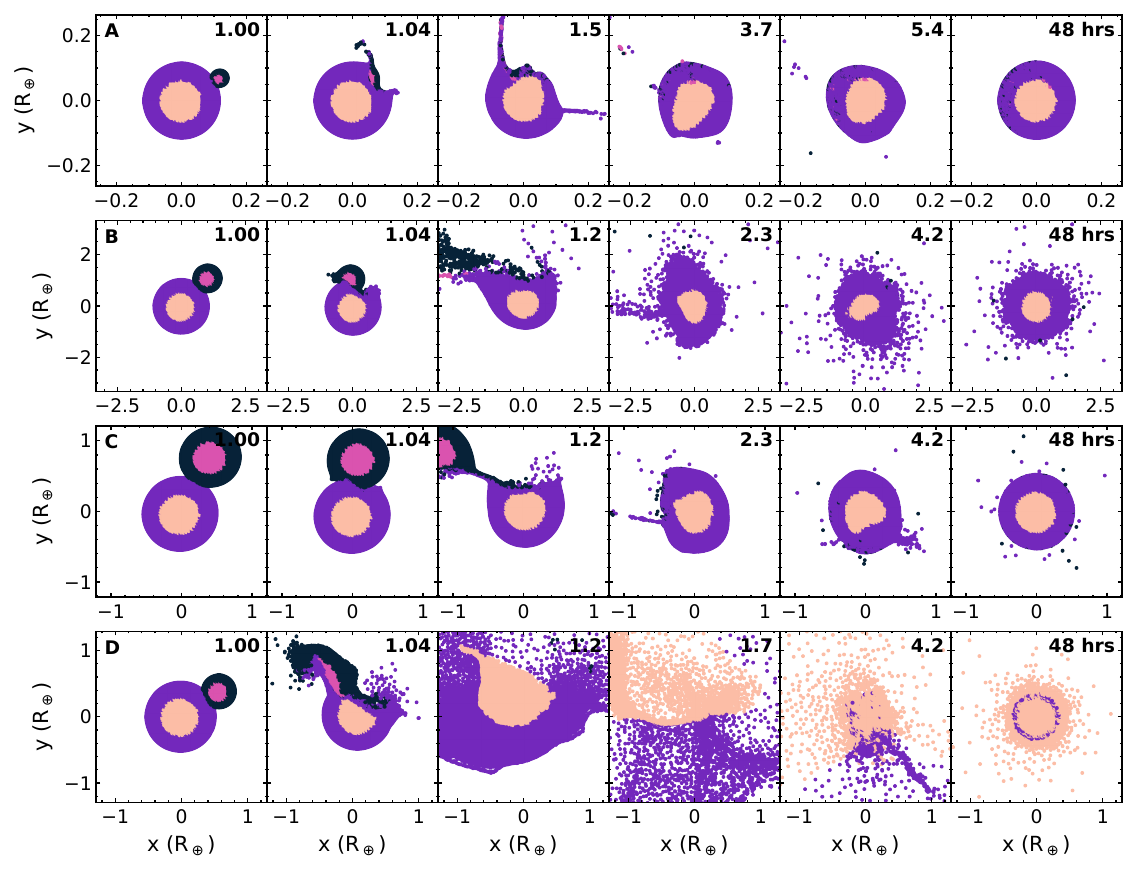}
\caption{{\bf Progression of four example simulated collisions showing materials.}
Materials and provenance of mass (core -- light/dark pink, mantle -- purple/dark blue -- for target/projectile) in the midplane as the same four example collisions shown in Figures \ref{f:collseq} and \ref{f:collseq2} progress.  The number in the upper right of each panel is the simulation time in hours. Plots show the SPH particle positions. 
A: an accretion collision with a 10$^{-3}$\,\mearth{} target and a mass ratio of 10$^{-2}$ at 3\,$v_\mathrm{esc}$ and an impact angle of 30$^\circ$ ($b = 0.5$). %cM1e-3p0mr1e-2v3.00b0.500N1e6Fo70S303Fe30S193. 
B: a mildly erosive collision with a 1\,\mearth{} target and a mass ratio of 0.1 at 5\,$v_\mathrm{esc}$ and an impact angle of 45$^\circ$ ($b = 0.707$). %cM1p0mr1e-1v5.00b0.707N1e6Fo70S303FeSi30S181. 
C: a hit-and-run collision with a 0.1\,\mearth{} target and a mass ratio of 0.5 at 4\,$v_\mathrm{esc}$ and an impact angle of 60$^\circ$ ($b = 0.87$). %cM1e-1p0mr5e-1v4.00b0.870N1e6Fo70S303FeSi30S182.
D: an erosive collision with a 0.1\,\mearth{} target and a mass ratio of 0.1 at 12\,$v_\mathrm{esc}$ and an impact angle of 30$^\circ$ ($b = 0.5$). %cM1e-1p0mr1e-1v12.00b0.500N1e6Fo70S303FeSi30S182.
Note a different scale is used for each collision.\label{f:collseqmat}}
\end{figure*}

\begin{figure*}%[!hbt]
\centering\includegraphics[width=0.94\textwidth]{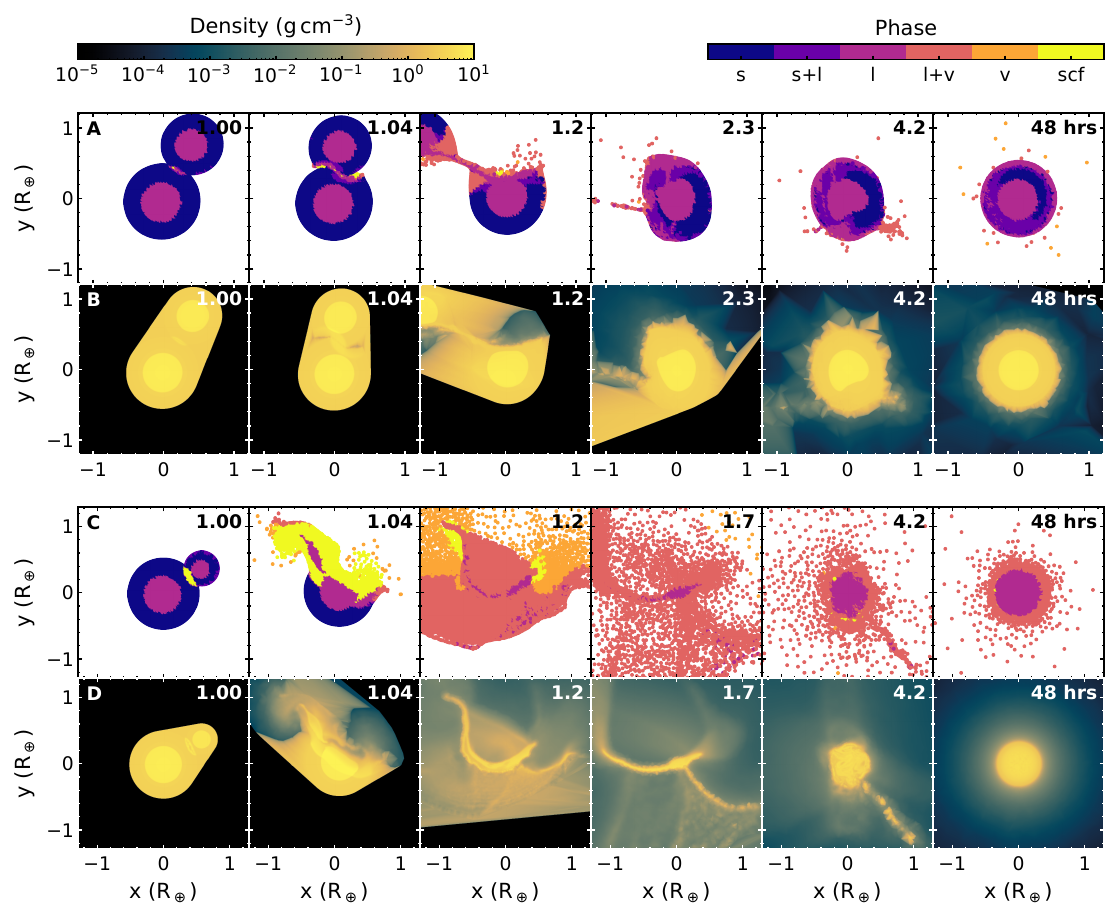}
\caption{{\bf Progression of two more example simulated collisions.}
Instantaneous phase (solid [s], mixed solid and liquid [s+l], liquid [l], mixed liquid and vapor [l+v], vapor [v], or super critical fluid [scf]) and density in the midplane as two example collisions progress.  The number in the upper right of each panel is the simulation time in hours. Phase panels show SPH particle positions, density panels show the fluid representation using interpolation. 
Rows A, B: a hit-and-run collision with a 0.1\,\mearth{} target and a mass ratio of 0.5 at 4\,$v_\mathrm{esc}$ and an impact angle of 60$^\circ$ ($b = 0.87$). %cM1e-1p0mr5e-1v4.00b0.870N1e6Fo70S303FeSi30S182.
Rows C, D: an erosive collision with a 0.1\,\mearth{} target and a mass ratio of 0.1 at 12\,$v_\mathrm{esc}$ and an impact angle of 30$^\circ$ ($b = 0.5$). %cM1e-1p0mr1e-1v12.00b0.500N1e6Fo70S303FeSi30S182.
Note a different scale is used for each collision. Corresponding materials plots are shown in Figure \ref{f:collseqmat}. \label{f:collseq2}}
\end{figure*}

Figure \ref{f:collseq2} shows two further examples of collisions, and \ref{f:collseqmat} shows the corresponding materials for the collisions shown in Figures \ref{f:collseq} and \ref{f:collseq2}.


\section{Vapor fraction}

\begin{figure*}
\centering\includegraphics[width=\textwidth]{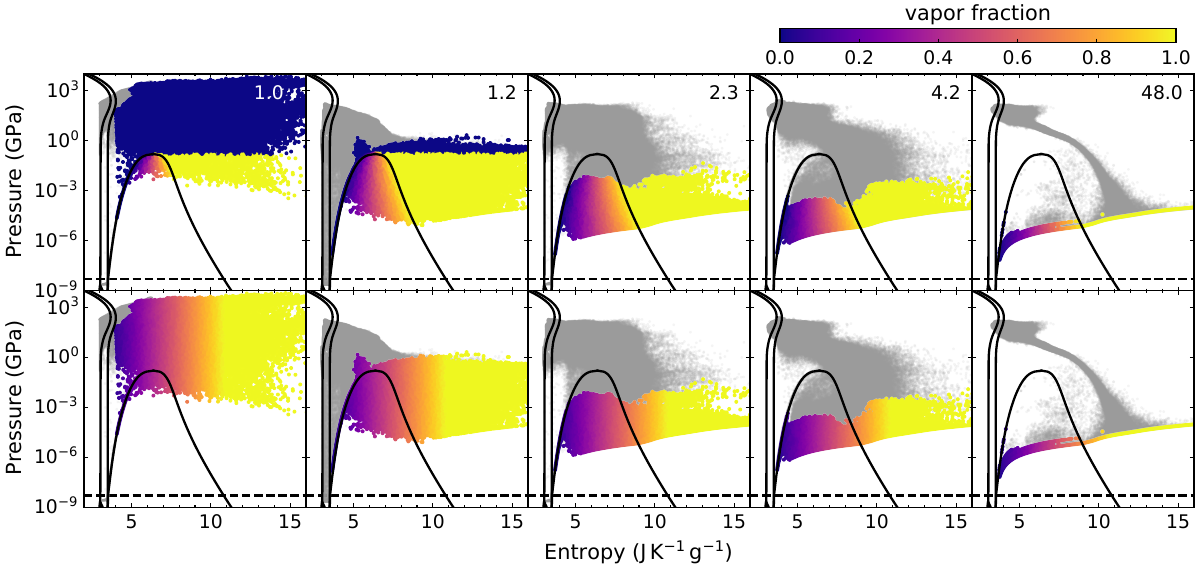}
\caption{{\bf The vapor fraction of ejecta calculated after decompression is generally lower than at the instantaneous pressure.}
Pressure vs entropy phase diagrams at five times during a collision with a 1\,\mearth{} target and a mass ratio of 0.1 at 5\,$v_\mathrm{esc}$ and an impact angle of 45$^\circ$ ($b = 0.707$) -- the second collision shown in Figures \ref{f:collseq} and \ref{f:collseqmat}. Individual SPH particles are plotted as grey dots. The upper row has ejecta particles coloured by instantaneous vapor fraction (supercritical fluid is considered as 0\% vapor), the lower row has ejecta particles coloured by vapor fraction calculated after isentropic decompression to 5.2\,Pa. The curve below which particles cannot drop visible in the lower half of the plots for the 3rd column onwards is due to the SPH density floor/maximum smoothing length.\label{f:phasedecompress}}
\end{figure*}

An example of the difference between instantaneous vapor fractions and vapor fractions after isentropic decompression to the triple point is shown in Figure \ref{f:phasedecompress}.

\section{Modified specific energy}\label{a:Qp}

The modified specific impact energy \citep{Leinhardt12} is defined as,
\begin{equation}
    Q'_\mathrm{R} = \frac{\mu_\alpha}{\mu}  Q_\mathrm{R} = \frac{1}{2} \mu_\alpha \frac{v^2}{M_\mathrm{tot}},
    \label{e:Qp_def}
\end{equation}
where $\mu$ is the reduced mass, $Q_\mathrm{R}$ is the specific impact energy, $v$ is the impact velocity, $M_\mathrm{tot}$ is the total mass, and $\mu_\alpha$ is a modified reduced mass that accounts for the interacting portion of the projectile's mass, $\alpha$ (see \citealp{Leinhardt12} and appendix \ref{a:alpha}):
\begin{equation}
    \mu_\alpha = \frac{\alpha M_\mathrm{proj} M_\mathrm{targ}}{\alpha M_\mathrm{proj} + M_\mathrm{targ}},
\end{equation}
where $M_\mathrm{targ}$ is the mass of the target body and $M_\mathrm{proj} = \gamma M_\mathrm{targ}$ is the mass of the projectile, and $\gamma$ is the mass ratio.

\section{Interacting mass}\label{a:alpha}

\citet{Leinhardt12} define the interacting mass, $\alpha$, as:
\begin{equation}
    \alpha = \frac{3rl^2 - l^3}{4 r^3},
    \label{e:alpha}
\end{equation}
where $r$ is the radius of the projectile and $l = (R+r)(1-b)$ is the projected length of the projectile that overlaps the target perpendicular to the velocity vector (see figure 2 in \citealp{Leinhardt12}), and $R$ is the radius of the target.

Assuming the average density of the projectile and target are equal, the ratio of the radii is just the cube root of the mass ratio, and we can write,
\begin{equation}
    r = \gamma^\frac{1}{3} R,
    \label{e:radratio}
\end{equation}
and,
\begin{equation}
    l = (1+\gamma^\frac{1}{3}) R (1-b).
    \label{e:l}
\end{equation}
Substituting equations \ref{e:radratio} and \ref{e:l} into equation \ref{e:alpha} we can derive an expression for $\alpha$ in terms of just $\gamma$ and $b$ (assuming the densities of target and projectile are the same),
\begin{equation}
    \alpha = \frac{\left[ 3 \gamma^\frac{1}{3} - \left(1+\gamma^\frac{1}{3}\right)(1-b)\right] \left(1+\gamma^\frac{1}{3}  \right)^2 (1-b)^2}{4 \gamma}.
\end{equation}

\section{Ejected mass}

\begin{figure}
\centering\includegraphics[width=0.7\columnwidth]{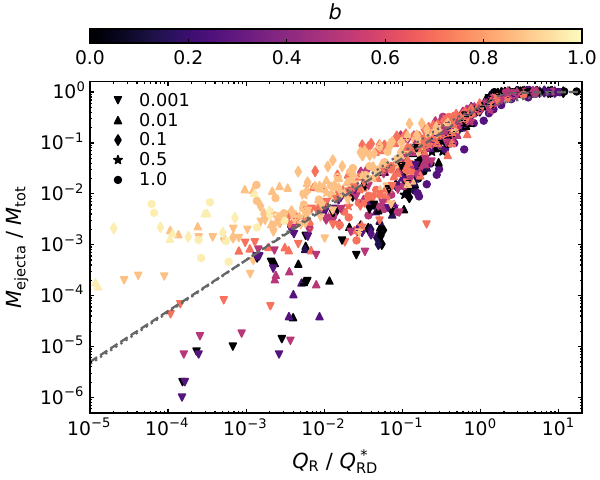}
\caption{{\bf The fractional mass of ejecta scales with the specific energy of the collision.}
Mass of ejecta as a fraction of total colliding mass as a function of normalised specific impact energy ($Q_\mathrm{R}/Q^*_\mathrm{RD}$). The shape of symbols corresponds to the mass ratio of the collision and the colour corresponds to the impact parameter. The dashed and dotted grey lines are the models from Leinhardt \& Stewart \citep{Leinhardt12} and Genda et al. \citep{Genda17_erosion} respectively. 
\label{f:Q_Mejecta}}
\end{figure}

The mass of ejecta from planetary collisions has received little previous attention, most studies have focused instead on the mass of the largest remnant(s). \citet{Genda17_erosion} examined the mass of ejecta from collisions involving large planetesimals (30--300\,km in radius). \citet{Leinhardt12} and \citet{Carter18} also investigated this regime, though their focus was the largest remnant mass rather than the ejecta mass. There are many collisions in our dataset that explore a similar regime of relatively low ejecta mass ($M_\mathrm{ejecta} < 0.1 M_\mathrm{tot}$), albeit with significantly larger targets ($>$600\,km radius).

Figure \ref{f:Q_Mejecta} shows the fractional mass ejected from collisions as a function of the normalised specific impact energy (see definition in \citealp{Leinhardt12}). Unsurprisingly, lower energy collisions generally result in less ejected mass. In general, higher impact parameter causes greater ejecta mass across the range of low ejecta mass collisions. Our results show good agreement with the predictions from \citet{Leinhardt12} and \citet{Genda17_erosion}.

\section{Ejecta vapor fraction}

\begin{figure*}
\centering\includegraphics[width=\textwidth]{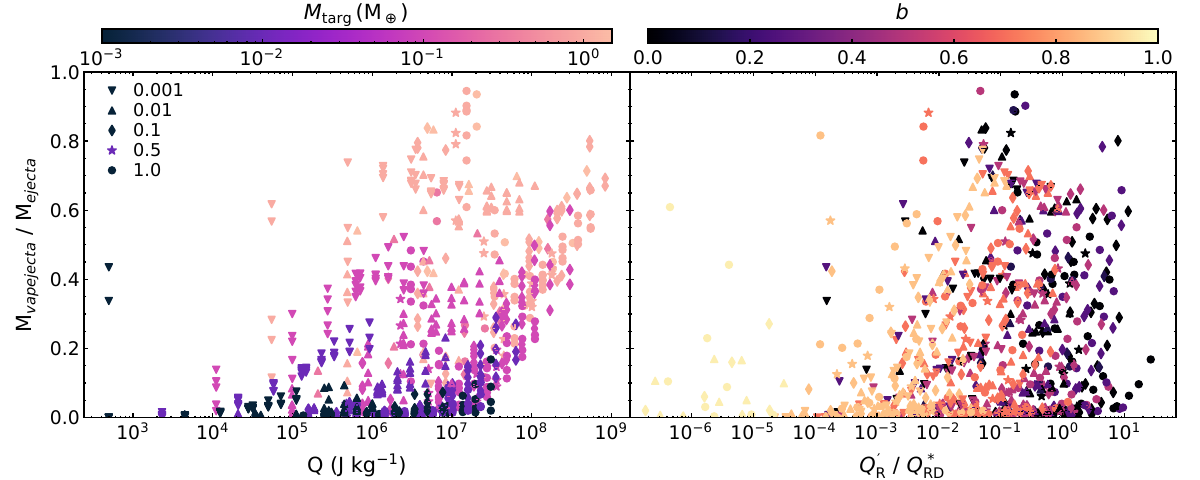}
\caption{{{\bf The vapor fraction of ejecta shows no clear trend with impact parameters.}
Mass of vaporized ejecta as a fraction of total colliding mass as a function of modified specific impact energy ($Q'_\mathrm{R}$). The shape of symbols corresponds to the mass ratio of the collision and the colour corresponds to the impact parameter. The dashed grey line is a power law fit to the data.}\label{f:Q_ejvapfrac_noncorr}}
\end{figure*}

{The ejecta vapor fraction (the vaporised ejecta mass as a fraction of the total ejecta mass) is plotted as a function of both specific energy and normalised specific energy in Figure \ref{f:Q_ejvapfrac_noncorr}.}

\section{Vaporized ejecta as a function of modified specific impact energy}

\begin{figure}
\centering\includegraphics[width=0.7\columnwidth]{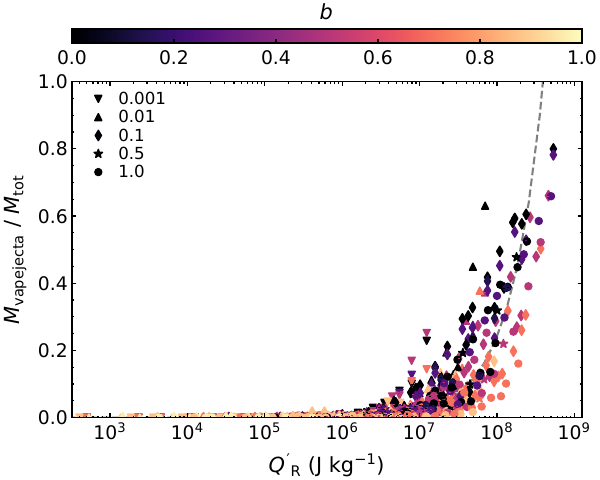}
\caption{{\bf The fractional mass of vaporized ejecta scales with modified specific impact energy.}
Mass of vaporized ejecta as a fraction of total colliding mass as a function of modified specific impact energy ($Q'_\mathrm{R}$). The shape of symbols corresponds to the mass ratio of the collision and the colour corresponds to the impact parameter. The dashed grey line is a power law fit to the data.\label{f:Q_ejvapfrac}}
\end{figure}

The relation between the fractional mass of vaporized ejecta and the modified specific impact energy of the collision is shown in Figure \ref{f:Q_ejvapfrac}.

\section{Scaling relation fit parameters}

The fit parameters for the vaporized ejecta mass relation obtained from our bootstrap fitting procedure are given in Table \ref{t:fitparams}.

\begin{table}
\caption{{\bf Vaporized ejecta mass scaling relation fit parameters.} Parameters from the bootstrap fitting for Fig. \ref{f:Mvap_K}. The data were fit in log space, thus the amplitude in equation \ref{e:vapejecta_K} is 10 to the power of the intercept and the power law index is the gradient. \label{t:fitparams}}
\begin{tabular}{lr}
\hline
\multicolumn{2}{c}{intercept (amplitude)} \\
mean & $-5.921$ \\
standard deviation \hspace{3cm} & $0.334$ \\
median & $-5.915$ \\
minimum & $-7.278$ \\
maximum & $-4.629$ \\
\hline
\multicolumn{2}{c}{gradient (index)} \\
mean & $1.006$ \\
standard deviation & $0.011$ \\
median & $1.006$ \\
minimum & $0.964$ \\
maximum & $1.051$ \\
\hline
\end{tabular}
    
\vskip4pt

\end{table}

\clearpage

\section{Full simulation parameters and results}

Full details of the simulation parameters and the derived results are given in Table \ref{t:fullsimlist}.

\footnotesize
%

\,

\clearpage

\end{document}